\documentclass[aps,prd,preprint,superscriptaddress,tightenlines,nofootinbib]{revtex4}

\usepackage{graphicx}
\usepackage{dcolumn}
\usepackage{bm}
\usepackage{epsfig}

\begin{document}


\newcommand{\ee}{e$^+$e$^-$}
\newcommand{\ff}{f$_{2}$(1525)}
\newcommand{\bb}{$b \overline{b}$}
\newcommand{\cc}{$c \overline{c}$}
\newcommand{\sbs}{$s \overline{s}$}
\newcommand{\uu}{$u \overline{u}$}
\newcommand{\dd}{$d \overline{d}$}
\newcommand{\qq}{$q \overline{q}$}
\newcommand{\suo}{\rm{\mbox{$\epsilon_{b}$}}}
\newcommand{\loro}{\rm{\mbox{$\epsilon_{c}$}}}
\newcommand{\kos}{\ifmmode \mathrm{K^{0}_{S}} \else K$^{0}_{\mathrm S} $ \fi}
\newcommand{\kol}{\ifmmode \mathrm{K^{0}_{L}} \else K$^{0}_{\mathrm L} $ \fi}
\newcommand{\ko}{\ifmmode {\mathrm K^{0}} \else K$^{0} $ \fi}

\def\tpc{three-particle correlation}
\def\twopc{two-particle correlation}
\def\ksks{K$^0_S$K$^0_S$}
\def\ee{e$^+$e$^-$}
\def\ff{f$_{2}$(1525)}

\preprint{CLEO CONF 03-05}   
\preprint{EPS-253}      

\title{Observation of $\eta_{c}^{\prime}$ Production in $\gamma\gamma$ Fusion
at CLEO~}
\thanks{Submitted to the
International Europhysics Conference on High Energy Physics,
July 2003, Aachen}

%
\author{J.~Ernst}
\affiliation{State University of New York at Albany, Albany, New York 12222}
\author{H.~Severini}
\author{P.~Skubic}
\affiliation{University of Oklahoma, Norman, Oklahoma 73019}
\author{S.A.~Dytman}
\author{J.A.~Mueller}
\author{S.~Nam}
\author{V.~Savinov}
\affiliation{University of Pittsburgh, Pittsburgh, Pennsylvania 15260}
\author{G.~S.~Huang}
\author{J.~Lee}
\author{D.~H.~Miller}
\author{V.~Pavlunin}
\author{B.~Sanghi}
\author{E.~I.~Shibata}
\author{I.~P.~J.~Shipsey}
\affiliation{Purdue University, West Lafayette, Indiana 47907}
% ...
\collaboration{CLEO Collaboration} 
\noaffiliation



\date{\today}

\begin{abstract}

We report on the observation of the
$\eta_{c}^{\prime}$(2$^{1}S_{0}$),
the radial excitation of $\eta_{c}$(1$^{1}S_{0}$)
ground state of charmonium, in the two-photon fusion reaction
$\gamma\gamma \rightarrow \eta_{c}^{\prime}
\rightarrow K_{S}^{0}K^{\pm}\pi^{\mp}$
in 13.4 fb$^{-1}$ of CLEO II/II.V data
and 9.2 fb$^{-1}$ of CLEO III data.
The data have been analyzed to extract the $\eta_{c}^{\prime}$ resonance
parameters.
\end{abstract}
\maketitle

\section{Introduction}

Knowledge of the hyperfine (spin singlet--triplet) splitting
is important for the understanding of the spin--spin interaction
in quarkonia. The existing experimental
knowledge is limited to the 117$\pm$2 MeV
splitting between $J/\psi$(1$^{3}S_{1}$) and
$\eta_{c}$(1$^{1}S_{0}$)\cite{pdg2002}.
It is important to find out the magnitude of the corresponding splitting
between the radial excitations
$\psi^{\prime}$(2$^{3}S_{1}$) and  $\eta_{c}^{\prime}$(2$^{1}S_{0}$), which
samples more confinement-dominated part of the $c\bar{c}$ potential whose
spin dependence is open to speculation.
The proximity of $\psi^{\prime}$ and  $\eta_{c}^{\prime}$
to the $D\bar{D}$ breakup threshold also brings in channel--coupling effects.
Several searches for  $\eta_{c}^{\prime}$ have been reported
in the literature\cite{cb}--\cite{l3}.

The Crystal Ball Collaboration at SLAC reported the identification
of  $\eta_{c}^{\prime}$ in the radiative decay of  $\psi^{\prime}$\cite{cb}.
Structure due to a purported $\gamma$ of energy 91$\pm$5 MeV was
interpreted as the signal for $M(\eta_{c}^{\prime})=3594\pm$ 5 MeV,
and a 95\% confidence level upper limit on its total width
was set as 8 MeV. The E760 experiment at Fermilab made an unsuccessful
search for  $\eta_{c}^{\prime}$ in the reaction
$\bar{p}p~\rightarrow~\eta_{c}^{\prime}~\rightarrow \gamma\gamma$
in the mass region 3591--3621 MeV with $\sim$6 pb$^{-1}$
of invested luminosity\cite{e760}. They set a 90\% confidence upper
limit for
$B(\eta_{c}^{\prime}\rightarrow\bar{p}p)\times\Gamma_{\gamma\gamma}(\eta_{c}^{\prime})$ of $\approx$0.7 eV
for assumed total widths between 5 and 10 MeV. The search was repeated
by the successor experiment E835 with the measurement with
$\sim$30 pb$^{-1}$ of luminosity. No evidence for  $\eta_{c}^{\prime}$
was found in the mass region 3575--3600 MeV, and a 90\% confidence
upper limit for
$B(\eta_{c}^{\prime}\rightarrow\bar{p}p)\times\Gamma_{\gamma\gamma}(\eta_{c}^{\prime})$ of  $\approx$0.7 eV was again set for assumed total
widths $\le$15 MeV\cite{e835}. It is worth noting that the corresponding
value for $\eta_{c}$ is
$B(\eta_{c}\rightarrow\bar{p}p)\times\Gamma_{\gamma\gamma}(\eta_{c})\approx$9
eV, {\it i.e.,} about an order of magnitude larger.

The next search for $\eta_{c}^{\prime}$ was done by DELPHI in the
$\gamma\gamma$ fusion reaction
$e^{+}e^{-} \rightarrow e^{+}e^{-} (\gamma\gamma) \rightarrow e^{+}e^{-}
(\eta_{c}^{\prime})$,
$\eta_{c}^{\prime}\rightarrow\rho^{0}\rho^{0},
K_{S}^{0}K\pi,K^{*0}K\pi,K_{S}^{0}K_{S}^{0}\pi\pi$ and $K^{+}K^{-}K^{+}K^{-}$.
No signal was observed in the mass range of 3500--3800 MeV, and a 90\% confidence upper limit
$\Gamma_{\gamma\gamma}(\eta_{c}^{\prime})/\Gamma_{\gamma\gamma}
(\eta_{c})\le0.34$ was set\cite{delphi} assuming
equal decay branching ratios for $\eta_{c}$ and $\eta_{c}^{\prime}$.
 In a similar search
L3 established a 95\% confidence limit of
$\Gamma_{\gamma\gamma}(\eta_{c}^{\prime})/\Gamma_{\gamma\gamma}
(\eta_{c})\le0.29$\cite{l3}.

Recently, the BELLE Collaboration has reported an
$\eta_{c}^{\prime}$ signal
with $>$$6\sigma$ significance in the decay of
$B\rightarrow~K\eta_{c}^{\prime}$, with
$\eta_{c}^{\prime}\rightarrow K_{S}^{0}K^{\pm}\pi^{\mp}$,
from which they determined
$M(\eta_{c}^{\prime})=3654\pm 6(stat)\pm 8(syst)$ MeV\cite{belle1},
and $\Gamma_{tot}(\eta_{c}^{\prime})$= 15 $^{+24}_{-15}$(stat) MeV.
The BELLE Collaboration also presented evidence
at the $3.4\sigma$ level
for an $\eta_{c}^{\prime}$ signal in the reaction
$e^{+}e^{-}\rightarrow (J/\psi) \eta_{c}^{\prime}$, with
$M(\eta_{c}^{\prime})=3622\pm 12(stat)$ MeV\cite{belle2}.

Theoretical predictions for the mass and width of the
$\eta_{c}^{\prime}$ meson
have been so far based on potential model calculations
and an analogy between the radial excitation  pairs
$(J/\psi)$/$\psi^{\prime}$ and $\eta_{c}$/$\eta_{c}^{\prime}$.
The predictions for masses range from
$M(\eta_{c}^{\prime})=3594$ MeV to $M(\eta_{c}^{\prime})=3629$ MeV,
and the predictions for the ratio of the two-photon partial widths,
$\Gamma_{\gamma\gamma}(\eta_{c}^{\prime})/
\Gamma_{\gamma\gamma}(\eta_{c})$ range from 0.38 to 0.77.
Thus, the current best experimental results for
both the  mass and two--photon partial width of the
$\eta_{c}^{\prime}$
are at
variance with theoretical predictions. This makes for great interest
in determining these parameters with precision.

At CLEO we have searched for the $\eta_{c}^{\prime}$ meson
in the final state $K_{S}^{0}K^{\pm}\pi^{\mp}$ via $\gamma\gamma$
fusion in  $e^{+}e^{-}$ data at the $\Upsilon$(4S) resonance
and its vicinity, using
13.4 fb$^{-1}$
of CLEO II/II.V data.
Having observed strong evidence for the $\eta_{c}^{\prime}$, we searched
for its corroboration in 9.2 fb$^{-1}$ of CLEO III data.
The observation of  $\eta_{c}$ in the same reaction
and decay channel was reported by CLEO earlier\cite{bran}.

Preliminary results of the present study were presented
earlier~\cite{cleo_conf}. The {\tt BaBar} Collaboration has also
recently presented
their preliminary results of the study of $\eta_{c}$ and  $\eta_{c}^{\prime}$
production in the same reaction~\cite{babar_conf}.

The layout of the paper is the following.
Section II describes
our data samples and event selection criteria.
Section III discusses our analysis procedures, while Section IV
presents our estimates of systematic uncertainties.
A summary of the results is given in Section V.



\section{Data Sample and Event Selections}

The  CLEO II/II.V data sample consists of 13.4 fb$^{-1}$ of
$e^{+}e^{-}$ luminosity.
Approximately one--third of the data were taken with the CLEO II
configuration of the detector, and two--thirds with the
CLEO II.V configuration.
The total luminosity of CLEO III data used in the current analysis is
9.2 fb$^{-1}$, $\sim 70\%$ of the CLEO II/II.V luminosity.
However, this newer data have the advantage of the superior
hadron identification, as described below.

The detector components most useful for this study were the
concentric tracking devices for measurements of
charged particles, operating in a
1.5 T superconducting solenoid. For CLEO II\cite{kubota}, this tracking
system consisted of a 6--layer straw tube chamber, a 10--layer precision
drift chamber, and a 51--layer main drift chamber. The main drift chamber
also provided measurements of the specific ionization loss, dE/dx, used
for particle identification.
For CLEO II.V,
the innermost chamber
was removed and replaced with a three--layer silicon
vertex detector\cite{hill}; in addition the gas
in the main drift chamber was changed from a 50--50 mixture
of argon--ethane to 60--40 helium--propane.

For CLEO III\cite{cleo3}, the entire charged particle tracking system
was replaced.
The new tracking
system consists of four layers of double-sided silicon strip detector,
surrounded by a new, 47-layer drift
chamber with a smaller outer
diameter\cite{dr3}.  Careful design optimization led to the
momentum resolution of this tracking system comparable to that
of CLEOII/II.V.
The CLEOII time-of-flight system was eliminated and in the space
created by it and the smaller drift chamber a ring imaging Cerenkov
detector (RICH)\cite{rich} was installed.

The RICH distinguishes charged kaons from pions over 80\% of the solid angle, which is smaller than hadron identification with dE/dx.
For charged tracks with momenta below 2 GeV/c
(the momentum range of most of our kaon candidates)
the RICH identifies kaons with efficiency greater than 81\% while having
less than 2\% probability that a pion is misidentified as a kaon.

In addition, the trigger for CLEOIII was improved in its
flexibility, efficiency, and redundancy.  Given that it relies on
fewer types of devices with better design than in CLEOII, the trigger
is also easier to reliably simulate.


The Monte Carlo simulation of CLEO detector
response was based upon {\tt GEANT}.  Simulated events were processed in
the same manner as the data to determine the $K_{S}^{0}K^{\pm}\pi^{\mp}$
detection efficiency.
Efficiencies obtained
in this fashion are listed for major selection criteria for both CLEO II/II.V
and CLEO III
in Table~\ref{tb:effi3}.
The overall detection efficiency for
$\eta_{c}^{\prime}$ is 13.8\%(CLEO II) and 11.6\%(CLEO III).

\begin{table*}
\begin{center}
\caption{Efficiencies for detecting the $\eta_c$ and
$\eta_{c}^{\prime}$ signals in \%.
Efficiencies for asterisked cuts were obtained for events which
have passed all other cuts.}
\label{tb:effi3}
\begin{tabular}{|c||c|c||c|c| }
\hline
& \multicolumn{2}{c||}{CLEO II/II.V} &  \multicolumn{2}{c|}{CLEO III} \\
\hline
& $\eta_c$ & $\eta_c^{\prime}$ & $\eta_c$ & $\eta_c^{\prime}$ \\
\hline
4 reconstructed tracks & 35 & 40 & 34 & 37 \\
\hline
$K_{S}$ identification & 77 & 79 & 61 & 64  \\
\hline
$K$/$\pi$ particle ID~~$^{*}$ & 90 & 93 & 75 & 78 \\
\hline
$P_{T} <$ 0.6 GeV/c~~$^{*}$ & 93 & 95 & 84 & 84 \\
\hline
Small $E_{neut}$~~$^{*}$ & 84 & 83 & 82 & 80 \\
\hline
Trigger~~$^{*}$ & 69  & 73 & 99 & 99 \\
\hline
\hline
Overall & 10.0 & 13.8 & 9.8 & 11.6\\
\hline
\end{tabular} \\
\end{center}
\end{table*}

The data sample used contained four charged particles, with at least one
$K_{S}^{0}$ candidate.
The $K_{S}^{0}$ vertex was reconstructed from its decay to
$\pi^{+}\pi^{-}$ and was required to be displaced from
the $e^{+}e^{-}$
interaction point. The mass of $K_{S}^{0}$ candidate was also required
to be  within 0.490 $<~ M(\pi\pi)~<$ 0.505 GeV/c$^2$
(0.488 to 0.508 GeV/c$^2$ for CLEO III).
Furthermore, the $K_{S}^{0}$ momentum vector was required to point
back to the interaction point.
The efficiency of these criteria, as determined from our
simulations, was
77$\%$
for $\eta_{c}$
and
79$\%$
for $\eta_{c}^{\prime}$ .
As shown in Table I, the CLEOIII efficiencies were somewhat smaller
than those of CLEO II/II.V.

The two charged particles other than those from  $K_{S}^{0}$ decay
were examined for the hypotheses
of being either kaons or pions.  In CLEO II/II.V,  this was done
%
 by using the dE/dx and TOF information (when available), and by applying
appropriate probability criteria.
The efficiency of this particle identification, again determined
from our simulations, was
90$\%$
for the $\eta_{c}$ and
93$\%$
for the $\eta_{c}^{\prime}$.

For CLEO III, all events were used in which the charged kaon
candidate is identified as a $K$ by
the RICH detector. When the $K$ candidate was not identifiable by the RICH
(mostly when it is outside the RICH fiducial
volume), it was identified by dE/dx measurements from the
drift chamber.  When its
momentum lay between 1 and 2 GeV/c, however, it was difficult to distinguish
$K$ from a charged pion using dE/dx and we did not use those $K$ candidates.

Other criteria used included using specific trigger requirements,
requiring a small transverse momentum of
the $K_{S}^{0}K^{\pm}\pi^{\mp}$ system ($P_{T}$$<$0.6 GeV/c), and
requiring the sum of neutral energy not associated with the
charged particles be minimal,
(E$_{neut}$$<$0.2 GeV for CLEOII/II.V; $<$0.3 GeV for CLEO III).
%

\begin{figure}[hbt]
\label{fg:fits}
\begin{center}
\includegraphics*[width=4.8in]{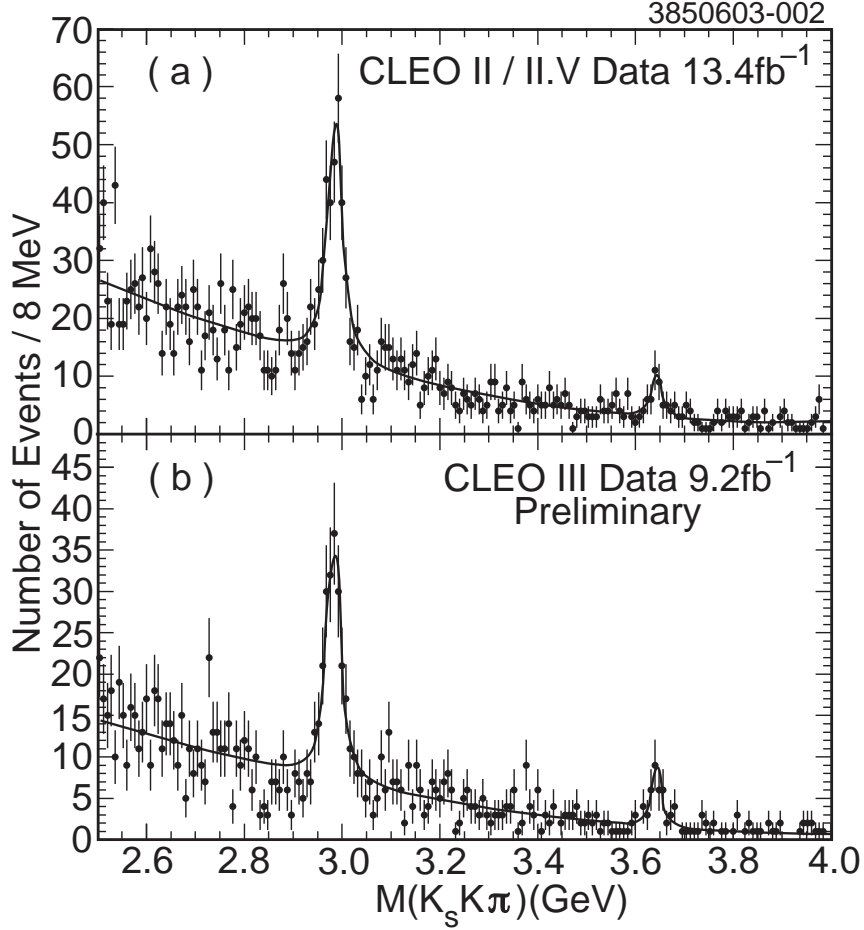}
\end{center}
\caption{$K_{S}^{0}K^{\pm}\pi^{\mp}$ invariant mass in the reaction
$\gamma\gamma\rightarrow K_{S}^{0}K^{\pm}\pi^{\mp}$
from the (a) the CLEO II/II.V data and (b) the CLEO III data. The curves in the
figures are fit results using a second order polynomial for the background.
For CLEO II/II.V two Breit--Wigners are used for the
$\eta_{c}$ and $\eta_{c}^{\prime}$ enhancements; for CLEO III
the fit shown uses Breit--Wigners convolved with Gaussians.}
\end{figure}

\section{Analysis}

The  $K_{S}^{0} K^{\pm} \pi^{\mp}$ invariant mass
plot using our final event selection for CLEO II/II.V
is shown in Fig.~1a.
Enhancements at masses $\sim$2985 and $\sim$3643 MeV
are visible.
Henceforth we call these
$\eta_{c}$ and   $\eta_{c}^{\prime}$ enhancements, respectively.

In order to extract numerical results from these data we
have made maximum likelihood fit to
this
spectrum using a second order polynomial background and two
Breit--Wigners, with their parameters kept free.  This fit
is overlaid in the Figure.
As presented in Table~\ref{tb:results}, the fit gives $36^{+15}_{-11}$
events at a mass of
$M(\eta_{c}^{\prime}) = (3642.7 \pm 4.1(stat.))$ MeV.
A fit using a Breit-Wiegner convolved with a double Gaussian
representing the resolution function extracted from Monte Carlo for each
%
%
of the signals is in good agreement with the fit using Breit-Wigners
only.
Note that we assumed that the potential interference between the $\eta_{c}$($\eta_{c}^{\prime}$) signal and the continuum two-photon
production of $K_{S}K\pi$ is negligible when we equate the peaks of
the distribution with the masses of the $\eta_{c}^{(\prime)}$.  The same assumption was
made for the rate measurements.

\begin{table}[hbt]
\caption{The results of the analysis.  The
yields of events and the masses are from the fits to the data.
The significance of the enhancements are based on changes of
likelihood as described in the text.  The CLEO III results are preliminary.}
\label{tb:results}
\begin{center}
\begin{tabular}{|c|c|c||c|c|} \hline
& \multicolumn{2}{c||}{CLEO II/II.V} &  \multicolumn{2}{c|}{CLEO III ({\em preliminary})} \\
\hline
   &  $\eta_{c}$ &  $\eta_{c}^{\prime}$ &
$\eta_{c}$ &  $\eta_{c}^{\prime}$ \\ \hline
Yield (events) & 287~$\pm$~28 &36 $^{+15}_{-11}$ & 203$\pm$22& 29$\pm$8\\
Mass (MeV) &  2984.7~$\pm$~2.1 &3642.7~$\pm$~4.1 & 2982$\pm$2& 3642.5$\pm$3.6\\
%
%
significance & 15.9$\sigma$ & 5.0$\sigma$ & 14.2$\sigma$& 5.7$\sigma$\\ \hline
\end{tabular}
\end{center}
\end{table}

The ``significance levels'' of the enhancements in this Table
were obtained as
$\sigma\equiv$$\sqrt{-2ln(L_{0}/L_{max})}$,
 where $L_{max}$ is the maximum likelihood returned by the fit
described above, and  $L_{0}$ is the likelihood returned by the fit
with either no $\eta_{c}$ or no $\eta_{c}^{\prime}$ resonance.
The search for  $\eta_{c}^{\prime}$ resonance was made in the mass
region 3580--3700 MeV.
Since the mass and width are not {\em a priori} known,
a more appropriate way of estimating the signal significance
is to determine the
probability that the counts for the background alone could
statistically fluctuate to the level of the counts of signal
plus background. It is found that the significance levels
determined in this manner is smaller than those listed in
Table~\ref{tb:results} by about 0.5$\sigma$.

Fig.~1(b) shows the preliminary $K_s^0K^{\pm}\pi^{\mp}$ invariant mass
distribution from CLEO III after all
criteria were applied.
It shows clear
enhancements at the $\eta_c$ mass as well as at the mass at which the
CLEO II/II.V data indicated the $\eta_{c}^{\prime}$.
This is strong confirmation that the  $\eta_{c}^{\prime}$
signal we observed in the CLEO II/II.V data is real and not due to statistical
fluctuation or some other artifact.

We fit the mass spectrum with second order polynomial
representing the background under the $\eta_{c}^{\prime}$ candidate peak and
a Breit-Wigner
function convolved with a double Gaussian representing the instrumental
resolution of 5.7 MeV (96\%) and 40 MeV (4\%)
as determined from our Monte Carlo simulation.
We have also used power--law background and
exponential background shapes and averaging over all three background shapes,
we obtain preliminary results of (29$\pm$8) events in the peak at a
mass of (3642.5$\pm$3.6(stat)) MeV. These fit results are also shown
in Table~\ref{tb:results}.
This measurement of the $\eta_{c}^{\prime}$ mass is completely
consistent with that obtained from the CLEO II/II.V data.
The statistical significance of the $\eta_{c}^{\prime}$
signal is determined to be  5.7$\sigma$, using the likelihood method
described above.  Note that in this case, this significance is the appropriate
quantity since we search for the $\eta_{c}^{\prime}$ at a particular
mass, namely that determined in the CLEO II/II.V analysis.


Photon--photon fusion is expected to populate positive charge
conjugation resonances mainly when the photons are almost real,
{\it i.e.,} when the transverse momenta of both of them, and therefore
of the sum of final state particles is small.
In order to examine whether the
observed $\eta_{c}^{\prime}$ peaks are
due primarily to two--photon events,
we examined the production of $\eta_{c}^{\prime}$ in two
subregions of transverse momentum: 0~$<$~$P_{T}$~$<$~0.2 GeV/c
and 0.2~$<$~$P_{T}$~$<$~0.6 GeV/c.
Table~\ref{tb:pt} shows the results of this study.
The CLEO II/II.V and CLEO III results
are statistically consistent with the expectations from our
two--photon Monte--Carlo simulations.

\begin{table}[hbt]
\caption{The transverse momentum dependence of production.}
\label{tb:pt}
\begin{center}
\begin{tabular}{|c|c|c|} \hline
$P_{T}$ & 0--0.2 GeV/c & 0.2--0.6 GeV/c \\
\hline
2 $\gamma$ Expectations & 74\% & 26\% \\
\hline
CLEO II/II.V &  $(55\pm 18)$\% &  $(45\pm 18)$\% \\
\hline
CLEO III & $(62\pm 19)$\% & $(38\pm 19)$\% \\
\hline
\end{tabular}
\end{center}
\end{table}

Also of interest is the two-photon partial width of the $\eta_{c}^{\prime}$
as compared to that of the $\eta_{c}$.
The ratio of the number of events of $\eta_{c}^{\prime}$ and $\eta_{c}$ is:

\begin{equation}
{N(\eta_{c}^{\prime}) \over N(\eta_{c})} =
{ \Gamma_{\gamma \gamma}(\eta_{c}^{\prime})
\times B(\eta_{c}^{\prime} \to K_{S}K\pi)
\over \Gamma_{\gamma \gamma}(\eta_{c}) \times B(\eta_{c}
\to K_{S}K\pi)}
\times{\Phi(m_{\eta_{c}^{\prime}}) \over \Phi(m_{\eta_{c}})}
\times{ \epsilon(\eta_{c}^{\prime}) \over \epsilon(\eta_{c})} \,~~~~~.
\end{equation}

Here, for the CLEO II/II.V analysis:

$N(\eta_{c}^{\prime})/N(\eta_{c}) = 0.125 ^{+0.054}_{-0.041}$ is the ratio
of $\eta_{c}^{\prime}$ and  $\eta_{c}$ events from Table~\ref{tb:results}

$\Phi(m_{\eta_{c}^{\prime}})/\Phi(m_{\eta_{c}}) = 0.42$ is the ratio of
the two--photon fluxes at the $\eta_{c}^{\prime}$ and $\eta_{c}$
masses,
as
calculated using the known two--photon flux function\cite{pdg2002}.
This ratio has a systematic uncertainty of $\pm$2\%.

${\epsilon (\eta_{c}^{\prime})/\epsilon(\eta_{c})} = 1.38$
is the ratio of efficiencies as calculated from Monte Carlo
simulation, and is shown in Table~\ref{tb:effi3}.

Using these values in the equation above, we obtain

\begin{equation}
R(\eta_{c}^{\prime}/\eta_{c})\equiv{\Gamma_{\gamma \gamma}(\eta_{c}^{\prime}) \times B(\eta_{c}^{\prime} \to
K_{S}K\pi) \over \Gamma_{\gamma \gamma}(\eta_{c}) \times
B(\eta_{c} \to K_{S}K\pi)} =  0.22^{+0.09}_{-0.07}(stat) \, .
\end{equation}


Performing this same calculation for the CLEO III data,
we obtain $R(\eta_{c}^{\prime}/\eta_{c})= 0.29\pm 0.09(stat)$,
which is preliminary but consistent with the CLEO II/II.V result.

\section{Systematic Uncertainties}

The systematic uncertainties in the CLEO III results are still being
investigated; therefore we only discuss those for the CLEO II/II.V
analysis below.

\subsection{Mass of the $\eta_{c}^{\prime}$ meson}

We estimate systematic uncertainties
on our mass determinations with the following considerations.

$\bullet$ It is difficult to estimate the contributions to systematic
biases which arise from different choices of event selection when
statistical uncertainties
 dominate. However, the statistical errors in the $\eta_{c}$
peak are much smaller than those of the  $\eta_{c}^{\prime}$ peak,
and we consider the extreme variation in $\eta_{c}$ mass
in the different event selections we have attempted as providing
an upper limit of 1 MeV as the systematic uncertainty
due to event selections.

$\bullet$ We have made a Monte Carlo study of how the masses of
 $\eta_{c}$ and  $\eta_{c}^{\prime}$ shift between the input
values and those obtained after the detector simulation.
For an input mass and width of
$M$=2980 MeV, $\Gamma_{tot}$=27 MeV, the shift,
$\Delta M(\eta_{c})$, is 0.9$\pm$0.6 MeV.
With an input mass and width of  $M$=3640 MeV, $\Gamma_{tot}$=15 MeV,
the shift, $\Delta M(\eta_{c}^{\prime})$, is 0.7$\pm$0.3 MeV.
We conclude that the systematic error due to this source
is  $\pm$1.0 MeV.

$\bullet$ We have also tested how the masses
of the particles that can be reconstructed compare with their
known masses. We find that the
reconstructed mass $M(K_{S}^{0})$=497.69$\pm$0.06 MeV. The
PDG\cite{pdg2002}
value is $M(K_{S}^{0})$=497.67$\pm$0.03 MeV. This small deviation,
$\Delta M(K_{S}^{0})$=0.02$\pm$0.07 MeV, is consistent with the conclusions
reached for the mass of the $J/\psi$ meson, namely
$\Delta M(J/\psi)$=0.4$\pm$0.4 MeV, which was measured separately.
We therefore conclude
that the absolute calibration of our mass scale has an uncertainty
of $\le$1.0 MeV.

$\bullet$ It is possible that the choice of background shape effects the
mass determination. From the fits made with polynomial,
 power--law, and exponential backgrounds we find that
$M(\eta_{c})$ varies by  $\le$0.7 MeV, and
$M(\eta_{c}^{\prime})$ varies by $\le$0.1 MeV. We therefore
estimate that this source contributes  $\le$ 1.0 MeV to the
systematic uncertainty in masses of $\eta_{c}$ and $\eta_{c}^{\prime}$.

Adding the above contributions in quadrature
we determine the systematic uncertainty in the mass of
 $\eta_{c}^{\prime}$ to be  $\le$2 MeV.
We note, however, that  $\eta_{c}$ mass obtained
in the present study (Table II) differs from our earlier published
result\cite{bran} by $\sim$4 MeV.
Provided this
difference, we assign the systematic uncertainty in our
measurement as $\pm$4 MeV, at present. Studies are being made to
understand the mass difference, and we expect that systematic uncertainties
at the level of $\pm$2 MeV will be achieved in the final results.

\subsection{$R(\eta_{c}^{\prime}/\eta_{c})$}

This value of $R$ cited earlier was obtained
using second order polynomial function for the background.
The major source of systematic uncertainty in the ratio $R$ is found
to be the choice of background shape. The values in case of
power--law and exponential backgrounds is essentially the same,
$R=0.13^{+0.05}_{-0.04}(stat)$ and
 $R=0.14^{+0.06}_{-0.05}(stat)$.
As our final result for $R$ we quote the average of the extreme values
and assign the difference  as systematic error. The systematic error
due to efficiency ratio was $\sim$2\%, which is negligible.

Thus,

\begin{equation}
R(\eta_{c}^{\prime}/\eta_{c})\equiv{\Gamma_{\gamma \gamma}(\eta_{c}^{\prime}) \times B(\eta_{c}^{\prime} \to
K_{S}K\pi) \over \Gamma_{\gamma \gamma}(\eta_{c}) \times
B(\eta_{c} \to K_{S}K\pi)}
=  0.17^{+0.07}_{-0.06}(stat)~\pm0.04(syst) \, .
\end{equation}

\section{Summary}

We have analyzed 13.4 fb$^{-1}$
of $e^{+}e^{-}$ data of CLEO II/II.V and 9.2
fb$^{-1}$ data  of CLEO III at $\Upsilon$(4S) and its vicinity
for resonances in the reaction
$e^{+}e^{-} \rightarrow e^{+}e^{-} \gamma\gamma \rightarrow e^{+}e^{-}
\eta_{c}^{\prime}\rightarrow e^{+}e^{-} (K_{S}^{0}K^{\pm}\pi^{\mp})$.

In the $K_{S}^{0} K^{\pm} \pi^{\mp}$ invariant mass plot resulting from
the CLEO II/II.V data,
we see an excess of events at mass of $\sim$3643
MeV with the significance level of over 4$\sigma$
in addition to $\eta_{c}$.
We attribute this excess
to the excitation of the  $\eta_{c}^{\prime}$  resonance.
Observation of an excess at the same mass
in the CLEO III data
confirms the discovery of $\eta_{c}^{\prime}$ in the CLEO II/II.V data.


Our results for the $\eta_{c}^{\prime}$ meson
are as follows.

The value of its  mass has been found to be:
\begin{equation}
M(\eta_{c}^{\prime})_{\rm CLEOII}=(3642.7 \pm 4.1 (stat)\pm 4.0(syst))~ {\rm MeV}
\, ,
\end{equation}
\begin{equation}
M(\eta_{c}^{\prime})_{\rm CLEOIII}=(3642.5\pm 3.6(stat))~ {\rm MeV (preliminary)}  \, ,
\end{equation}
thus,
\begin{equation}
\Delta M\equiv M(\psi^{\prime})-M(\eta_{c}^{\prime})=
(43 \pm 4 (stat) \pm 4(syst))~ {\rm MeV} \, .
\end{equation}

We also determine the ratio
\begin{equation}
R(\eta_{c}^{\prime}/\eta_{c})_{\rm CLEOII}\equiv{\Gamma_{\gamma \gamma}(\eta_{c}^{\prime}) \times B(\eta_{c}^{\prime} \to
K_{S}K\pi) \over \Gamma_{\gamma \gamma}(\eta_{c}) \times
B(\eta_{c} \to K_{S}K\pi)} =
0.17^{+0.07}_{-0.06}(stat) \pm 0.04(syst) \, .
\end{equation}



We gratefully acknowledge the effort of the CESR staff
in providing us with
excellent luminosity and running conditions.
This work was supported by
the National Science Foundation,
the U.S. Department of Energy,
the Research Corporation,
and the
Texas Advanced Research Program.

\end{document}